\documentstyle[11pt]{article}
\def\hybrid{\topmargin 0pt      \oddsidemargin 0pt
	\headheight 0pt \headsep 0pt
	\textheight 9in         
	\textwidth 6.25in       
	\marginparwidth .875in
	\parskip 5pt plus 1pt   \jot = 1.5ex}

\catcode`\@=11
\def\marginnote#1{}
\newcount\hour
\newcount\minute
\newtoks\amorpm
\hour=\time\divide\hour by60
\minute=\time{\multiply\hour by60 \global\advance\minute by-\hour}
\edef\standardtime{{\ifnum\hour<12 \global\amorpm={am}%
	\else\global\amorpm={pm}\advance\hour by-12 \fi
	\ifnum\hour=0 \hour=12 \fi
	\number\hour:\ifnum\minute<10 0\fi\number\minute\the\amorpm}}
\edef\militarytime{\number\hour:\ifnum\minute<10 0\fi\number\minute}

\def\draftlabel#1{{\@bsphack\if@filesw {\let\thepage\relax
   \xdef\@gtempa{\write\@auxout{\string
      \newlabel{#1}{{\@currentlabel}{\thepage}}}}}\@gtempa
   \if@nobreak \ifvmode\nobreak\fi\fi\fi\@esphack}
	\gdef\@eqnlabel{#1}}
\def\@eqnlabel{}
\def\@vacuum{}
\def\draftmarginnote#1{\marginpar{\raggedright\scriptsize\tt#1}}

\def\draft{\oddsidemargin -.5truein
	\def\@oddfoot{\sl preliminary draft \hfil
	\rm\thepage\hfil\sl\today\quad\militarytime}
	\let\@evenfoot\@oddfoot \overfullrule 3pt
	\let\label=\draftlabel
	\let\marginnote=\draftmarginnote
   \def\@eqnnum{(\theequation)\rlap{\kern\marginparsep\tt\@eqnlabel}%
\global\let\@eqnlabel\@vacuum}  }

\def\numberbysection{\@addtoreset{equation}{section}
	\def\theequation{\thesection.\arabic{equation}}}

\def\underline#1{\relax\ifmmode\@@underline#1\else
	$\@@underline{\hbox{#1}}$\relax\fi}

\def\titlepage{\@restonecolfalse\if@twocolumn\@restonecoltrue\onecolumn
     \else \newpage \fi \thispagestyle{empty}\c@page\z@
	\def\thefootnote{\fnsymbol{footnote}} }

\def\endtitlepage{\if@restonecol\twocolumn \else  \fi
	\def\thefootnote{\arabic{footnote}}
	\setcounter{footnote}{0}}  
\catcode`@=12
\relax

\def\beq{\begin{equation}}
\def\eeq{\end{equation}}
\def\bea{\begin{eqnarray}}
\def\eea{\end{eqnarray}}
\def\bar{\overline}

\def\nn{\nonumber}

\def\TB{\widetilde B}

\def\CD{{\cal D}}
\def\CF{{\cal F}}

\def\nn{\nonumber}

\relax
\numberbysection
\hybrid
\begin{document}
\begin{titlepage}
\begin{center}
 \hfill    PAR--LPTHE 95--55  \\
 [.5in]
{\large\bf  FIELD ANTI-FIELD DUALITY,
P-FORM GAUGE FIELDS\\ AND\\
TOPOLOGICAL  FIELD THEORIES}\\[.5in]
        {\bf   Laurent Baulieu }\footnote{email address:
baulieu@lpthe.jussieu.fr} \\
    	   {\it LPTHE\/}\\
       \it  Universit\'e Pierre et Marie Curie - PARIS VI\\
       \it Universit\'e Denis Diderot - Paris VII\\
Laboratoire associ\'e No. 280 au CNRS
 \footnote{ Boite 126, Tour 16, 1$^{\it er}$ \'etage,
        4 place Jussieu,
        F-75252 Paris CEDEX 05, FRANCE}
\\
and\\
RIMS, Kyoto University\\
 Kyoto, 606-01 Japan

\end{center}

\vskip .5in

\begin{quotation}
\noindent{\bf Abstract }    We construct  a  framework which  unifyies in dual
pairs  the fields and
anti-fields of the Batalin and Vilkovisky quantization method. We consider
  gauge theories of p-forms coupled to
 Yang-Mills fields. Our  algorithm   generates many
topological models of the Chern-Simon type or of the Donaldson-Witten type.
Some of these models   can undergo a  partial   breaking of their topological
symmetries.
\end{quotation}
 \end{titlepage}
 \newpage

 \section{Introduction} The   field and anti-field formalism  which has been
developed by Batalin and Vilkovisky (B-V) attracts more and more attention
\cite{BV}\cite{henneaux}.
 Anti-fields were first introduced in the context of the renormalization
of the  Lagrangian of Yang-Mills  gauge theories
 as the sources of the local operators representing  the BRST variations of the
propagating fields,  with a  master equation which controls their
renormalization \cite{zinn}.  Later on, after earlier developments in
Hamiltonian formalism, Batalin and Vilkovisky have built a Lagrangian
formalism,
where one associates for each one of the classical and
ghost fields,  collectively denoted as
$\phi$,  an anti-field $\phi^*$ \cite{BV}.

In the construction of Batalin and Vilkovisky, the local action   which
determines the quantum theory is a local functional $S[\phi,
\phi^*]$   satisfying the B-V    graded master equation
\begin{equation} {\delta S \over \delta \phi} {\delta S \over \delta \phi^*}
\pm {\delta S \over \delta \phi} {\delta  S \over \delta \phi^*} = 0
\end{equation}

The definition of the graded differential BRST operator $s$    is
\begin{equation}\label{defs}
s\phi = {\delta S[\phi, \phi^*] \over \delta \phi ^*} \quad  \quad
s\phi^* = {\delta S [\phi, \phi^*] \over \delta \phi}
\end{equation}
 $S[\phi,
\phi^*]$  has ghost number zero by assumption. Thus,  if the field  $\phi$  has
ghost number $g$, its anti-field  $\phi^*$
 has ghost number $-g-1$.

The B-V equation  takes its most transparent form if one defines the
following graded bracket acting in the space of functionals of $\phi$ and
$\phi^*$
\def\be{\begin{equation}} \def\ee{\end{equation}}
\be
\{X,Y\}={\delta X \over \delta \phi} {\delta Y \over \delta \phi^*}
\pm {\delta Y \over \delta \phi} {\delta  X\over \delta \phi^*}
\ee
where $X$ and $Y$ are functionals. Then one has
\be
s= \{S,\ \}
\ee
and the B-V equation is
\be
 sS= \{S,S \}=0
\ee
With this notation,   the nilpotency property
\be
s^2\phi=s^2\phi^*=0
\ee
is an obvious consequence of the  B-V master equation and of the
graded Jaccobi identity of the bracket $\{\ ,\  \}$.

 Reciprocally, if there is a way to
 define  directly an operation $s$ acting on a set of fields and anti-fields
$\phi$   and    $\phi^*$ with the property $s^2=0$, and if a
local functional  $S[\phi,\phi^*]$ exists with $sS=0$,  then one has $s= \{S,\
\}
$,
 that is, eq.  (\ref{defs}) holds, and $S[\phi,\phi^*]$ can be identified with
the B-V action. This      property  will be used systematically in this paper:
we
will construct from an algebraic principle the BRST symmetry for large classes
of
gauge theories  of forms coupled to a Yang-Mills field and deduce only
afterwards the B-V action  $  S[\phi, \phi^*]$ and the classical action $
S[\phi, 0]$.

One of the great advantages
 of the B-V formalism is of permitting   a consistent  quantization of
actions invariant under gauge symmetries whose transformation laws
close only modulo some of the equations of motion. Examples of such symmetries
include   supergravities, models with  non-abelian   form gauge fields with
degree larger than one
 and  open
string field theory  \cite{bbs} \cite{boc}.

 From the point of view of quantum field theory,
fields and anti-fields seem to play dissymmetric roles.  In our present
understanding, anti-fields are not quantum fields: they are to be eliminated
from the action through the choice of a   local gauge function
$Z (\phi)$ with ghost number minus one by mean of the constraint
  \begin{equation}
\phi^* = {\delta Z  [\phi] \over \delta \phi}
\end{equation}

  With apropriate choices of
$Z $, $S(\phi, \phi^*={\delta Z  [\phi] \over \delta \phi})$ becomes
   a consistently gauge fixed action   which
contains generally higher-order ghost interactions. Formal proofs which are
based on the nilpotency of the
operation $s$ before the elimination of anti-fields  show   that
physical quantities    do not  depend on the choice  of the functional
$Z(\phi)$
\cite {BV}  \cite {henneaux}. The    classical action
 $S_0[\phi]=S[\phi, 0]$      is
invariant under the restricted part $s_0$ of the BRST symmetry operator
defined by  $s_0 \phi=s \phi\big| _{\phi^*=0}$.       In general, the
nilpotency of $s_0$ is broken by terms proportional to  the  equations of
motion, a property which originates in the fact that
the BRST variations  of the
anti-fields involve the  equations of motion of  $S_0[\phi]$.

In a previous attempt
 to incorporate the B-V  formalism in a geometrical picture,   some kind of
unification between fields and anti-fields has been shown to  exist in
particular cases
\cite{bbs} \cite{boc}.

 In this paper we will obtain a more general result.
By considering       the  gauge theories of forms,  including
Yang-Mills and scalar fields, we will show that a sort   of
duality   exists  between the fields and the anti-fields of the B-V
quantization
in the framework of    a  beautiful algebraic structure.
More precisely,
 given a $p$-form gauge field in D-dimensional space, valued in a given
Lie group representation, we will show that it has a natural   ``dual''
companion which is a $D-1-p$ gauge field. The argument is
that  the anti-fields of $B_p$  and of its   ghosts and ghosts of ghosts
    are contained in an expansion which includes  negative ghost number
components  for
$B_{D-1-p}$,  and vice-versa. This implies in particular that
the natural companion of a Yang-Mills field in 4 dimensions is a   2-form gauge
field.
Furthermore, we  will show that quite simple algebraic formulae determine  the
  BRST equations
as constraints on curvatures and,   eventually, the   B-V
actions.

The formalism that we shall present generates in a systematic way  many
topological actions functions of p-form
gauge fields. They are generalizations  of the Chern-Simon  action  and/or
 of the  Donaldson-Witten  action \cite{wittendona}.  We get therefore
theories which are either defined from
classical Hamiltonians which vanish  up to gauge transformations,
or from  the gauge-fixing of classical
actions which are equivalent to   topological terms.

There is a simple explanation for this possibility  of unifying the  fields and
the anti-fields presented in this paper.
Indeed, the BRST formalism has to do with a superfield
formalism in a superspace $\{x^\mu, \theta\}$, where $\theta$ is a scalar
Grassman variable and $x^\mu$ are the ordinary real coordinates of the
D-dimensional space.
 Forms should
be expanded on monomials products of $dx^\mu$ and $d\theta$.
Since $dx^\mu$ is  odd, the ordinary form degree of any given form  can only
take
  integer values between 0 and $D$. On the other hand, $d\theta$ is a
commuting object, and we have the freedom to consider monomials of the type
$(d\theta)^g$ with no restriction on   the possible  values of $g$.
In
particular,  $g$ can be  a negative integer. Our proposal is  that  anti-fields
must be identified as forms
with a negative ghost number  (which should not be confused with the  antighost
 number).

To be more precise, let us consider     the tangent
plane defined above
 the point with local coordinates $(x^\mu, \theta=0$). One has the following
decomposition   for a p-form $\tilde B_p(x,\theta=0)$ living in this space
\begin{equation}
\label{decompo}
\tilde B_p(x,\theta=0) = \sum_{q=0}^{D} B_q^{p-q}(x)
\end{equation}
with
\begin{equation}
\label{dec}
B_q^{p-q}(x) = {1 \over q!}\  B_{\mu_1\cdots \mu_q}^{p-q}
(x)\  dx^{\mu_1}    {\wedge \cdots \wedge} dx^{\mu_q} \ (d\theta)^{p-q}
\end{equation}

The physical interpretation of this equation is that  $B_p^0(x)$ is a classical
$p$-form
gauge field  and that the fields
$B_q^{p-q}(x)$,
 with $0 \leq  g \leq p-1$, are the ghosts and ghosts of ghosts of
$B_p(x)$  \cite{cargese}.

The     fields  $B_q^{p-q}(x)$, for  $p-D \leq  p-q \leq  -1$, have a  negative
ghost
number, counted by the negative power of $(d\theta)^{p-q}$.
The aim of this paper is to identify these new fields   as the anti-fields of a
``dual''
($D-1-p$)-form gauge field and of its ghosts and ghosts of ghosts.
This is a consistent identification   because (i) the anti-field of a
field with ghost number $g$ has ghost number $-g-1$  and (ii)
forms with ordinary form degree
$q$ or $D-q$ contain  as many independent
Lorentz components in D-dimensional space.

Before   giving the details of our    general construction  and  some
examples of interest, we will briefly list  a few properties of $p$-form
gauge fields.

\section{Properties of p-form gauge fields}

Consider a classical $p$-form   gauge field
 valued in a given Lie Algebra $\cal G$

\begin{equation}
B_p(x)={1\over p!}
B_{\mu_1\cdots \mu_p}(x)
dx^{\mu_1} {\wedge\cdots\wedge}dx^{\mu _p}
\end{equation}

This form contains $C^p_D$
independent components in $D$-dimensional space.
It is expected  that   a gauge    field is massless and truly lives
in a $(D-2)$-dimensional space, that  is, in
the hyperplane transverse to its propagation.
Thus, by generalizing Feynman argument, one must   introduce
  ghosts and anti-ghosts to add up positive and negative degrees of freedom and
obtain a system of fields counting for an effective number of degrees of
freedom
equal to
$C^p_{D-2}$. This can be achieved by extracting the ghosts and antighosts from
the following expansion \cite{cargese}
 \begin{equation}
B_p\to \TB_p=
\sum^p_{g=0}
\sum^g_{q=0}
B^{g-q,q}_{p-g}
\end{equation}
The upper indices $g-q$ and $q$ are respectively the ghost number and the
anti-ghost number of the form
$B^{g-q,q}_{p-g}$,
which has ordinary form degree $p-g$.

Let us briefly justify this decomposition,
the goal of which is to substract unwanted
degrees of freedom.
One defines the total degree of a field as the sum of its usual
form degree and of its ghost and anti-ghost numbers.
In this sense, each
term of the series of fields  which defines  $\TB_p$ is a $p$-form.
 Moreover, one defines the
statistics of the field
$B^{q,g-q}_{\mu_1\cdots\mu_{p-g}}$
 as  even (resp.\ odd) if $g$ is even (resp.\ odd). This definition
would become  a tautology
  in the superspace notation of eq. (\ref{dec})
with an additional     $\bar\theta$
direction
to accommodate for the anti-ghost components.

With these definitions,   the number of propagating
``physical'' degree of freedom  $N_D(p)$ of the p-form gauge
field in $D$-dimensional space is obtained by adding  the  degrees of freedom
of each field occuring in the definition of $\TB_p$ with an algebraic
weight $1$   for the field components with even
    statistics and $-1$ for those  with odd  statistics. It is rather simple
to   find the following formula for $0\le p\le D-2$
\be    N_D(p)=\mathop\sum\limits^p_{g=0}(-)^g(1+g)C^g_{p-g}=C^p_{D-2}
 \nonumber
\ee
and
\be
 N_D(D-1)=N_D(D)
=0.
\ee
This is the wanted property which ensures that we have a system of fields
which amounts to a p-form    existing  in the transverse
plane  with $D-2$ dimensions.
Obviously $(D-1)$- and $D$-forms
carry no degree of freedom in $D-2$ dimension space,
which explains physically
the result $N_D(D-1)=N_D(D)=0 $.

In what follows, we will forget the antighost components, since the non
trivial sector of the BRST symmetry is only for
 the fields with no anti-ghost component, of the type $B^{g,0}_{p-g}$.
 The extension to the anti-ghost sector   of the formulae that we will
derive would be obvious by introducing Nakanishi-Lautrup type auxiliary
fields and  an anti-BRST operation, but it is not the subject of this paper.

We shall therefore focus on the possible ways of defining    the BRST operator
on the following object
\begin{equation}
{\hat B}_{p}=\sum^p_{g=0}
B^g_{p-g}.
\end{equation}
We will find equations which determine directly $s \hat B_{p}$, the remaining
obvious task being the decomposition in ghost number of these equations.

 In
\cite{cargese}, it was shown that, given a collection of forms
$B^{(i)}_{p_i}$,
the determination of the BRST operator can be cast into  the construction of
curvatures
\begin{equation}
{\hat G}^{(i)}_{p_i+1}
\equiv (d+s)
 {\hat B  }^{(i)}_{p_{i+1}}+
 {  K}^{(i)}_{p_{i+1}}(\hat B )
\end{equation}
upon which one puts "horizontality" constraints
\begin{equation}
\hat G^{(i)}_{p_{i+1}}={1 \over (p+1)!}\  G^{(i)}_{\mu_1\cdots \mu_{p+1}}
(x)\  dx^{\mu_1}    {\wedge \cdots \wedge }dx^{\mu_{p+1}}
\end{equation}
provided that
the form of the field polynomials $K^i_{p_{i+1}}$
is compatible with the Bianchi identity stemming from
\begin{equation}
s^2=0\quad sd+ds=0\Leftrightarrow
(s+d)^2=0.
\end{equation}

This construction can   be applied to   the ordinary Yang-Mills
theory  and  also   to less trivial examples such that
the theories of forms coupled  to  Yang-Mills fields through
a Chern-Simon term as in the Chapline-Manton symmetry \cite{chapline} as well
as
to theories invariant
under reparametrization and local supersymmetry
\cite{grimm}.   However, there are cases  for which this approach is not fully
satisfying,
because  it   provides  a BRST symmetry whose nilpotency is
broken by terms proportional to the gauge covariant equations of motions. Such
cases require that one uses the B-V formalism.

The content of the  next sections is thus to show  that  the
freedom of introducing in the expansion of a $p$-form gauge fields objects
with negative ghost numbers  permits one to reconciliate  the B-V approach
and the algebraic framework.

\section{Field anti-fields unification of   Yang-Mills  fields and (D-2)-forms}

Let us consider   $D$-dimensional space. The basic object that we must
introduce is a Yang-Mills field $A=A_\mu dx^\mu$ valued in a Lie algebra
${\cal G}$.

According to  the first section of this paper, one should consider the
following generalized
$\cal G$-valued one-form:
\begin{equation} \tilde A(x)=A^{1-D}_D+A^{2-D}_{D-1}+\cdots
+A^{-2}_3 +A^{-1}_2+A+c
\end{equation}

The field $c$ is   the  Faddeev-Popov ghost
of $A$. Since the anti-field of a field with ghost number
 $g$ has ghost number
$-g-1$  and since  $A^{-p+1}_p$ has the same number of Lorentz components as a
 form with ordinary form degree $(D-p)$, one can identify the fields with
negative ghost numbers
$A^{-1}_2,A^{-2}_3, A^{-3}_4,\dots ,A^{-D+1}_D$
 as
the anti-fields of a $\cal G$-valued $(D-2)$-form $B_{D-2}$, of its ghost
$B^{1}_{D-3}$ and of its ghosts  of ghosts $B^{2}_{D-4},\dots ,B^{D-2}_0$.

It is therefore natural to introduce another fundamental form which is  a
generalized
$\cal G$-valued (D-2)-form based on a  (D-2)-form gauge field
$B _{D-2}$  \begin{equation}
\tilde B_{D-2}(x)=B^{D-2}_0+B^{D-3}_1+\cdots +B^{1}_{D-3}+B _{D-2}+B^{-1}_{D-1}
+B^{-2}_D
\end{equation}
with  \begin{equation}
(B^{p}_{D-2-p})^*=A^{-p-1}_{p+2}\quad p\ge 0
\end{equation}
It is rewarding that   the objects with negative ghost numbers in
the expansion of the 2-form $\tilde B_{D-2}$, namely $B^{-1}_{D-1}$ and
 $B^{-2}_D$,  can be considered respectively as the anti-fields
of the Yang-Mills one form $A=\tilde A^0_1$ and of its Faddeev-Popov
  ghost $c=\tilde A^1_0$
\begin{equation}
(B^{-2}_D)^*=c\qquad (B^{-1}_{D-1})^*=A
\end{equation}
One has therefore   the following field anti-field   relations
between $\tilde A$ and $\tilde B_{D-2}$
\be
  (A^{1-p}_p)^*=B^{p-2}_{D-p}
\end{equation}
and
\be
A^{1-p}_p=(B^{p-2}_{D-p})^*
\ee
for $\quad 0  \leq p \leq D$. This relation is an involution, and we find it
apropriate  to
call it a duality relation.

We desire to find algebraic equations
 generalizing the algorithm of
\cite{cargese} which only involve $\tilde A$
and $\tilde B$ and determine the action of the possible BRST operators $s$ on
all fields and
anti-fields. For this purpose,
 we define
\begin{eqnarray}
D^{\tilde A}&=&d+[\tilde A,\quad   ]\nonumber \\
F^{\tilde A}&=&d\tilde A+\frac{1}{2}[\tilde A, \tilde A]
\end{eqnarray}
and
\begin{eqnarray}
\CD&=&s+  D^{\tilde A}= d+s+[\tilde A, \quad ]\nonumber \\
\CF&=&(s+d)\tilde A+\frac{1}{2}[\tilde A, \tilde A]
=s\tilde A+ F^{\tilde A}
\end{eqnarray}
The   relation
\begin{equation}
(s+d)^2=0
\end{equation}
  amounts to the equation
\begin{equation}
\label{bian}
\CD\CD=[\CF,\quad ]    \qquad \CD\CF=0
\end{equation}

Therefore, to determine consistently the action of $s$, that is with
$(s+d)^2=0$, we must
simply  put  constraints on the generalized curvatures $\CF $ and $\CD\tilde
B_{D-2}$ of  $\tilde A$
and $\tilde B_{D-2}$ which are compatible with eq. (\ref{bian}).

For generic values of the space   dimension, we have the solution
\begin{eqnarray}
\label{basics}\CF&=&0\nonumber \\
\CD\tilde B_{D-2}&=&0
\end{eqnarray}
that is
\begin{eqnarray}
\label{base}
-s\tilde A&=&F^{\tilde A}\nonumber \\
-s\tilde B_{D-2}&=& D^{\tilde A}\tilde B_{D-2}
\end{eqnarray}

The expansion in ghost number of $F^{\tilde A}$ is
\begin{eqnarray}
F^{\tilde A}&=& \frac{1}{2}[c,c]+D^Ac+(dA+\frac{1}{2}[A,A]+[A^{-1}_2,c])
\nonumber \\
 & &+(D^AA^{-1}_2+[c,A^{-1}_3])+\cdots+[c,A_D^{-D+1}]
\end{eqnarray}

Thus, eqs. (\ref{basics})  give the following expression for
 the nilpotent BRST transformations of all
fields and anti-fields
\begin{eqnarray}
sc &=& -\frac{1}{2}[c,c]
 \nn\\
sA &=& -D^A c     \nn\\
sA^{-1}_2&=&s(B_{D-2})^*=-F^A-[c,A^{-1}_2]
\nonumber \\
& \vdots & \nonumber \\
sA^{-D+1}_D&=&s(B^{D-2}_0)^*=-D^A
A^{-D+2}_{D-1}-[c,A^{-D+1}_D]-[A^{-1}_2,A^{-D+2}_{D-2}]
-\dots
 \nn
\end{eqnarray}
and
\begin{eqnarray}
\label{base1}
 sB^{D-2}_0&=&-[c,B^{-D-2}_0]\nonumber \\
sB^{D-3}_1&=&-[c,B^{D-3}_1]-D^AB^{D-2}_0\nonumber \\
sB^{D-4}_2&=&-[c,B^{D-4}_2]-D^AB^{D-3}_1-[A^{-1}_2,B^{0-2}_0]\nonumber \\
\vdots\nonumber \\
sB_{D-2}&=&-[c,B_{D-2}]-D^A B^1_{D-3}-[A^{-1}_2,B^2_{D-4}]-
\cdots +[A^{-D+3}_{D-2},B^{D-2}_0]\nonumber \\
sB^{-1}_{D-1}&=&s(A^*)=-[c,B^{-1}_{D-1}]-D^A B_{D-2}-\cdots\nonumber \\
sB^{-2}_{D }&=&s(c^*)=-[c,B^{-2}_{D}]-D^A B^{-1}_{D-1}-\cdots
\end{eqnarray}

If one sets equal to zero the anti-fields one gets the intuitive BRST
transformation of a 2-form gauge field $s_0  B_{D-2}=-[c,B_{D-2}]-D^A
B^1_{D-1}$. $s_0$ is  nilpotent only for $F^A=0$.

To find the  B-V action corresponding  to the symmetry defined in eqs.
(\ref{base1}),   we observe that
eqs. (\ref{base}) imply the following cocycle equation
\begin{equation} (d+s)T_r(\tilde F
\wedge \tilde B_{D-2})=0
\end{equation}
Thus the invariant  B-V action    is
 \begin{equation}
S[\tilde A,\tilde B_{D-2}]=\int T_r\left [\tilde B_{D-2} \wedge F^{\tilde
A}\right ]^O_{D}
\end{equation}
By expansion in   field components of the forms
$\tilde A$ and $ \tilde B_{D-2}$ , one gets
\begin{eqnarray}
S[\tilde A,\tilde B_{D-2}]=\int T_r \ (&& B_{D-2}\wedge F^A-\frac{1}{2}B^{-2}_D
[c,c] -B^{-1}_{D-1}D^Ac  \nonumber \\
&&+A^{-1}_{2}\wedge (-[c,B_{D-2}]-D^A B^1_{D-3}\nn\\
&&-[A^{-1}_2,B^2_{D-4}]-
\cdots +[A^{-D+3}_{D-2},B^{D-2}_0]) +\ldots
 )
\end{eqnarray}
By setting all anti-fields equal to zero, one can verify in particular that
the classical
action is a $BF$ action \cite{review}.

 One can add to
the B-V action
 $S[\tilde A,\tilde B_{D-2}]$  a gauge
invariant action $S_{cl}[A]=\int{\cal  L}_{cl}(A)$, for instance $\int
F^2_{\mu v}d^4x$  or a Chern-Simon action  (for odd values of the space
dimension D). Then one should replace the BRST symmetry defined in eqs.
(\ref{basics})  by
\begin{eqnarray}
\CF&=&0\nonumber \\ \CD\tilde B_{D-2}&=&\frac{\delta S_{cl}}{\delta A}
\end{eqnarray}
This modification is   apparently   spurious, since  the
equation of motion of the field $ B_{D-2}$ implies the vanishing curvature
equation $F^A=0$.  We will come back on this point in the last section
and see how the vanishing curvature condition could be mildened.

\section{Coupling to   p-form gauge fields: Chern-Simon
type actions}

Let us now introduce a $p$-form gauge field $X_p$ in addition to the
  dual pair  $(\tilde A, \tilde B_{D-2})$. The integer $p$ is such that
$0\le p \le D-1$. We must consider
 the following field anti-field decomposition comparable to
 eq. (\ref{decompo})
\begin{eqnarray}
\label{XX}
\tilde X_p =
 X^{p-D}_D+X^{p+1-D}_{D-1}+\cdots +
 X^{-1}_{p+1}+X_p+X^1_{p-1}+\cdots +X^p_0
\end{eqnarray}
To interpret the anti-fields occuring in eq. (\ref{XX}), we   introduce the
dual form $\tilde Y_{D-p-1}$, such that $\tilde Y_{D-p-1}^*=\tilde X_p$,
\begin{eqnarray}
\tilde Y_{D-1-p} =
 Y^{D-1-p}_0+Y^{D-2-p}_1+\cdots+Y^{1}_{D-p-2} +
 Y_{D-p-1}+Y^{-1}_{D-p}+\cdots +Y^{-1-p}_D
\end{eqnarray}
The BRST equations are defined by
\begin{eqnarray}
\label{con}
 {\cal F}
  &=& s\tilde A + d\tilde A + \frac{1}{2}[ \tilde A, \tilde A]=0 \nonumber \\
{\cal D}\tilde B_{D-2}
  &=& s\tilde B _{D-2}+D^{\tilde A}\tilde B_{D-2}
      = [\tilde X_p, \tilde Y_{D-1-p}] \nonumber \\
{\cal D}\tilde X_p
  &=& s\tilde X _p + D^{\tilde A}\tilde X _p =0 \nonumber \\
{\cal D}\tilde Y_{D-1-p}
  &=& s\tilde Y_{D-1-p} + D^{\tilde A}\tilde Y _{D-1-p} =0
\end{eqnarray}
The fact that these equations define    $s$
with $s^2=0$ is easily verified  from  the   identities obtained by
applying $  {\cal D}$ on both sides of eqs. (\ref{con}), using the
 equations
${\cal F} = {\cal D} {\cal D}  =0$ and ${\cal D\cal F}=0$.

The existence of a  B-V  equation follows from
the equation
\be
(s+d)\tilde{\cal L}_D = 0 \nonumber
\ee
with
\be
\tilde{\cal L}_D
=
 T_r(\tilde B_{D-2}\wedge   F^{\tilde A}
               +\tilde X _pD^{\tilde A}\tilde Y_{D-1-p})
\ee
The B-V action is thus
\begin{eqnarray}
S[\phi , \phi ^*] = \int \tilde {\cal L}^0_D &=&
        \int T_r \ [\ B_{D-2}\wedge F^A+X_p\wedge D^AY_{D-p-1}\nonumber \\
     & &+\sum_{q \not = 0}
       B^q _{D-2-q} \wedge  F^{\tilde A} |^{-q }_{q +2}\nonumber \\
     & &+\sum_{q \not= 0}
       X^q _{p-q} \wedge (D^{\tilde A} \tilde  Y_{D - p  -1}))|^{-q}_{D +q-p}
\ ]
\end{eqnarray}
  It is a simple exercise to  verify that the BRST transformations stemming
from this B-V action are identical to those following from the constraints
defined in
eqs. (\ref{con}).

 To understand the nature of the
model let us
 consider the   classical    action
\begin{equation}
S_{cl}[\phi  ]=S [\phi , \phi ^*=0]
=\int T_r\left(\ B_{D-2}\wedge F^A + X_p \wedge D^AY_{D-2}\right )
\end{equation}
$S_{cl}$   is invariant under the following gauge symmetry, obtained by
equating to zero all anti-fields  and by replacing in the BRST transformations
of
the classical fields
 the primary ghosts $c, B^1_{D-3}, X^1_{p-1}, Y^1_{D-p-2} $
by infinitesimal parameters $\epsilon, \epsilon_{D-3}, \epsilon_{p-1},
\epsilon_{D-p-2}$ \begin{eqnarray} \delta A &=& D^A\epsilon \nonumber \\
\delta B_{D-2}  &=& D^A\epsilon _{D-3}+[\epsilon, B_{D-2}]
     +[\epsilon_{p-1}, Y_{D-p-1}]+[X_p, \epsilon_{D-p-2}] \nonumber \\
\delta X_p &=& D^A\epsilon _{p-1}+[\epsilon , X_p] \nonumber \\
\delta Y_{D-1-p} &=& D^A\epsilon _{D-2-p}+[\epsilon , Y_{D-1-p}]
\end{eqnarray}
The classical equations of motion are
\begin{eqnarray}
F^A&=&0 \nonumber \\
D^AX_p&=&0 \nonumber \\
D^AY_{D-1-p}&=&0
\end{eqnarray}
The model is thus quite similar to the Chern-Simon theory.

 There is of course the possibility of adding other actions made
from  several pairs $(\tilde X_{p_i}, \tilde Y_{D - p_i -1})$, with all
possible values of $p_i$.
The B-V action is in this case
\begin{equation}
S = \int T_r\left[\tilde B_{D-2}  \wedge F^{\tilde A}
+ \sum_i \tilde X_{p_i}\wedge D^{\tilde A} \tilde  Y_{D - p_i -1}\right]^0_D
\end{equation}

\section{Topological actions stemming from  d-exact
Lagrangian}
The Yang-Mills topological BRST symmetry is based on the following
BRST transformations of the Yang-Mills fields
and of its Faddev-Popov ghost \cite {review}
\begin{eqnarray}
\label{top}
sA & = & -Dc + \Psi \nonumber \\
sc & = & -\frac 12 [c,c] + \Phi
\end{eqnarray}
$\Psi=\Psi_\mu dx^\mu$ is a one-form with ghost number one and
$\Phi$ is a commuting scalar ghost with ghost number two. We will show that
this type of symmetry enters naturally in the field anti-field  dual  framework
explained in this paper.

We first consider the case   $D \neq 4$.
 In addition to  the dual pair
$(\tilde A, \tilde B_{D-2})$, we introduce
  another $\cal G$-valued dual pair
$(\tilde X_{2}, \tilde Y_{D-3})$. The expansions in form components of $ \tilde
X_{2} $ and $ \tilde Y_{D-3} $ are similar to the expression given in   eq.
(\ref{XX}) and contain all possible fields and anti-fields compatible with the
form degrees $2$ and $D-3$.

We then define the nilpotent $s$-operation by the following constraints
compatible with Bianchi identities
\begin{eqnarray}
 {\CF}  & = & s\tilde A  + d\tilde A + \tilde A\tilde A = \tilde X_2
 \nonumber \\
\CD \tilde X_2 & = & s\tilde X_2  +  D^{\tilde A}\tilde X_2
 = 0 \nonumber \\
\CD \tilde B_{D-2} & = & s\tilde B_{D-2}  + D^{\tilde A}\tilde B_{D-2}
 =   [\tilde X_2,\tilde Y_{D-3}] \nonumber \\
\CD \tilde Y_{D-3} & = & s\tilde Y_{D-3} + D^{\tilde A}\tilde Y_{D-3}
 =0
\end{eqnarray}
The first equation gives the   BRST
topological symmetry defined in eq. (\ref{top}) with
$\Psi = \tilde X_1^1$
and $\Phi = \tilde X_0^2$.

Furthermore, the above constraints imply
\be
(s+d)\ T_r(\tilde B_{D-2}\wedge  ( F ^{\tilde A}+ \tilde{X}_2)
+ \tilde X_2  \wedge   D ^{\tilde A}\tilde Y_{D-3})=0
\ee
It follows that the B-V action of the system is
\begin{equation}
S(\phi, \phi^*) = \int \tilde T_r\left [B_{D-2}(  F ^{\tilde A}+ \tilde X_2)
+ \tilde X_2   D ^{\tilde A}\tilde Y_{D-3}\right]_D^0
\end{equation}
and the  classical action is
\begin{equation}
S_{cl}(\phi, \phi^*=0) = \int T_r\left (B_{D-2}\wedge (F^A+X_2) + X_2 \wedge
D^AY_{D-3}\right)
\end{equation}
The  field $X_2$ can be eliminated by
its equation of motion, with
\begin{equation}
S_{cl}(\phi,\phi^*=0) \sim \int T_r\left(D^AY_{D-3}\wedge F^A\right)
=\int d\ T_r\left(Y_{D-3}\wedge F^A\right)
\end{equation}
This shows that the B-V action, after the replacement  of the anti-fields by
antighosts  via the standard procedure \cite{henneaux}, should be the gauge
fixing of
a topological term
$\int d\ T_r Y_{D-3}\wedge F^A $.
We are thus considering theories of the Donaldson-Witten type. As an
example, for D=3, one gets the topological field theory based on the
Bogolmony equations  \cite{bs}.

 Let us consider now the exceptional case
$D=4$. In this important particular case,
  $\tilde B_{D-2}$
is a 2-form   and can be used as a substitute for
$\tilde X_2$.
The system becomes  simplest and one has
\begin{eqnarray}
\CF & = & s\tilde A + F^{\tilde A} = \tilde B_2 \nonumber \\
\CD \tilde B_2 & = & s\tilde B_2 + D^{\tilde A }\tilde B_2 = 0
\end{eqnarray}
One has   $(s+d)\ T_r[\ \tilde B_2\wedge ( F ^{\tilde A}+ \tilde{B}_2) ]=0$.
Thus the  B-V
action is
\begin{equation}
S[\phi, \phi^*] = \int T_r \left[
\tilde B_2 \wedge (\tilde F + \frac12 \tilde B_2)\right]^0_4
\end{equation}
and the classical action is
\begin{equation}
S[\phi,\phi^*= 0] = \int T_r  \ B_2 \wedge  (F + \frac12 B_2) \end{equation}
By eliminating the  field $B_2$ by its equation of motion,
one finds
$S[\phi, 0]\sim \int T_r F \wedge F$.
This coincides with the fact that  the 4-D Yang-Mills topological action is
the gauge fixing of the second Chern-class \cite{bs}.

To conclude this section, let us remark that we have  not  considered the
possibility  that $p=D$. This is indeed a very special case, since   in the
decomposition $\tilde X_D = \sum\limits_{p=0}^D X_{D-p}^p$
one finds only terms with positive ghost numbers. No anti-field occurs in
the decomposition of such a $D$-form.    One must therefore introduce a form
made of  all
the anti-fields
$X_{p}^{-1-p}$
of the fields $X_{D-p}^p$. Such a generalized differential
form deserves the name
of a
$-1$-form gauge field, since it has the following decomposition
\begin{equation}
\tilde Y_{-1}= X_0^{-1} + X_1^{-2} + \cdots +
X_D^{-1-D}
\end{equation}
The BRST equations are now
\begin{eqnarray}
\CF & = & 0 \nonumber \\
\tilde\CD \tilde B_{D-2} & = & [\tilde X_D , \tilde Y_{-1}] \nonumber \\
\tilde\CD \tilde X_D & = & 0 \nonumber \\
\tilde\CD \tilde Y_{-1} & = & 0
\end{eqnarray}
and  the B-V action  action is still
$\int T_r(\tilde F \wedge \tilde B_{D-2}
+ \tilde X_D \wedge  D ^{\tilde A} \tilde Y_{-1})_D^0$.
We see however that no equation of motion exists
for $\tilde X_D$,
which is in fact absent from the classical action.

 \section{The case of three-dimensional space}
It is worth mentioning the case of $D=3$
dimensions. In this case,  one can consider the components with negative ghost
number of
$\tilde A$ as the  anti-fields of its components with positive or zero  ghost
number. In this sense,  $\tilde A$  is a``self-dual'' potential
\begin{equation}
\tilde A = A_3^{-2} + A_2^{-1} + A + c
\end{equation}
with
\begin{equation}
A_3^{-2} = c^* \qquad\qquad A_2^{-1} = A^*
\end{equation}
The BRST symmetry is defined by
\begin{equation}
\tilde\CF = s\tilde A + d\tilde A + \tilde A \tilde A = 0
\end{equation}
and one has
\begin{equation}
(s + d)\ T_r\left(\tilde A d \tilde A + \frac 23 \tilde A\tilde A\tilde
A\right) =0
\end{equation}
Thus, the Batalin-Vilkovisky action is simply
\begin{eqnarray}
S(A, A^*) & = & \int T_r\left[\tilde A d \tilde A  + \frac23
\tilde A \tilde A \tilde A \right]_3^0 \nonumber \\
& = & \int T_r \left( AdA + \frac23 A^3 +A_2^{-1}Dc + A_3^{-2}cc\right)
\end{eqnarray}
which is the standart  result for the
B-V action  for   the Chern-Simon  theory including its invariance under the
ordinary Yang-Mills symmetry \footnote{ For interesting results about the
quantization of this action and the
  correspondance with the unification that we have found here, see
ref. \cite{singerCS}}.

  It is quite natural  to introduce   a  ${\cal
G}$-valued scalar
field $\tilde \varphi = \varphi + \varphi_1^{-1} +
\varphi_2^{-2}+\varphi_3^{-3}$
with its dual 2-form
$\tilde Y_2 = Y_3^{-1} + Y_2 + Y_1 ^1 + Y_0^2$. To do so we must relax the
condition that the Yang-Mills field is "self-dual". We introduce another ${\cal
G}$-valued one-form
$\tilde a$, distinct from the Yang-Mills one-form   $\tilde A$, such that
$\tilde a$ is the dual of  $\tilde A$.
The symmetry is   defined now as
\begin{eqnarray}
\label{truc}
\  \CF  & = & s\tilde A  + d \tilde A + \tilde A\tilde A = 0
 \nonumber \\
\CD \tilde a & = & s\tilde a  + D^{\tilde A}\tilde a = F^A
 + [\tilde Y_2,\tilde \varphi]
 \nonumber \\
\CD \tilde \varphi  & = & s\tilde \varphi   + D^{\tilde A}\tilde \varphi = 0
\nonumber \\
\CD \tilde Y_2 & = & s\tilde Y_2 + D^{\tilde A}\tilde Y_2 = 0
\end{eqnarray}
 The corresponding B-V action is
\begin{equation}
S[\phi,\phi^*] = \int T_r\left[A\wedge dA+ \frac23 A\wedge A\wedge A+
  \tilde a \wedge F^{\tilde A}
+ \tilde Y_2 \wedge D^{\tilde A} \tilde \varphi \right]_3^0
\end{equation}
It is instructive enough  to write the classical action
\begin{equation}
 \int T_r\left ( A\wedge dA+ \frac23 A\wedge A\wedge A+  a \wedge F ^A+
Y_2\wedge
D^A\varphi\right)
\end{equation}
This action is
interesting as a generalized Chern-Simon  type action
involving   couplings of the Yang-Mills to a scalar field and a 2-form gauge
field. The equations of motion are
$F^A=0$  as in the genuine Chern-Simon  theory and $D^A a= D^A \varphi=D^A
Y=0$. The BRST
symmetry operator $s_0$ for  the classical action is
\begin{eqnarray}
s_0A & = & - D^Ac \nonumber \\
s_0a & = & - D^Aa_0^1 - [ c, a] - [Y_1^1, \varphi] \nonumber \\
s_0\varphi & = & -[ c, \varphi] \nonumber \\
s_0Y_2 & = & -D^AY_1^1 - [c, Y_2]
\end{eqnarray}

The quantization and the gauge-fixing of this action would necessitates that
one uses    the
full symmetry stemming from eq. (\ref{truc}),
 including the anti-fields
with $sY_2 = -DY_1^1 -[c,Y_2] -[A_2^{-1},Y_0^2]$
and $sA_2^{-1} = F^A - [c, A_2^{-1}]$.

\section{Possible breaking of the topological invariance
toward the creation of physical excitations}

We have shown in the previous sections a rather general way to produce actions
which are of the topological type in the sense that they have vanishing
Hamiltonians (up to gauge transformations) or are of the Donaldson-Witten type.
In this section we intend to sketch possible scenarios which could  break at
least partially
the topological symmetries of these models and possibly provide  models with
physical excitations.

 From now on, we restrict to
$D=4$ dimensions. In a quite generic way,
we have been led to consider actions of the type
\begin{equation}
\label{last}
S_4=\int T_r\left(\tilde B_2 \wedge F^{\tilde A}+
\tilde \varphi D^{\tilde A}\tilde C_3\right)^0_4
\end{equation}
Here
$\tilde \varphi =
\varphi^4_{-4}+\varphi^3_{-3}+\varphi^2_{-2}+\varphi^1_{-1}+\varphi$
is a generalized $0$-form, and
$\tilde C_3=C^{-1}_4+C_3+C^1_2+C^2_1+C^3_0$
is its dual. The action (\ref{last}) determines  a theory with  a Yang-Mills
field coupled to a  scalar $\varphi $, a $2$-form $B_2$ and a $3$-form
$C_3$.

This model could be useful for the purpose of computing
mathematical quantities from the path integral point
of view. However, the addition of gauge invariant terms like  like
$F^2_{\mu \nu}$ and
$(D_\mu \varphi)^2$ seems   of no relevance,
since the equations of motion of $B_2$ and $c_3$
would imply $F_{\mu \nu}=0$ and
$D_\mu \varphi =0$.

There is a first possibility of getting out of this situation. It  consists in
freezing the Yang-Mills symmetry, while keeping all other local
symmetries. Indeed, everywhere in our formula,  we can      put
  $c=0$, provided one has also $sA=0$.  By doing  so, one can
      add to the Lagrangian  the  term   $A^2_\mu$, which  yields actions as in
Freedman-Townsend model \cite{Freedman}.  By   eliminating  the field $B_2$,
the  constraint $F^A=0$ arises. It can  be solved with $A$ equal to a pure
gauge,
which gives a Lagrangian term  $A^2_\mu=(g^{-1}\partial_{\mu v}g)^2$. One
gets a non-linear sigma model, with possible couplings to
$X_p$ and $Y_{D-1-p}$.

The second possibility is to introduce a symmetry breaking mechanism,
 by adding to the action
from an ordinary Higgs potential  \be V(\varphi) = -\mu^2\varphi^2 + \lambda
\varphi ^4
\ee
The
symmetry of the action is
\begin{eqnarray}
\CF&=&s\tilde A+F^{\tilde A}=0\nonumber \\
\CD\tilde B_2&=&s\tilde B_2+D^{\tilde A}\tilde B_2=[\tilde C_3,\varphi]
\nonumber \\
\CD\tilde C_3&=&s\tilde C_3+D^{\tilde A}\tilde C_3=
{{\delta ^*V} \over {\delta \varphi}}\nonumber \\
\CD\tilde  \varphi&=&s\tilde \varphi+D^{\tilde
A}\tilde \varphi=0 \end{eqnarray}
If the potential $V$ is chosen such that  $\langle \varphi \rangle \not= 0$, we
get from these equations \begin{equation}
sB_2=[\langle \varphi \rangle, c^1_2]+\cdots
\end{equation}
This implies that we can gauge fix to zero certain  components of the 2-form
gauge field
$B_2$   along
  group directions. This might  relax the
constraints that the Yang-Mills curvature vanishes along these directions.
   Our
claim is  thus that one can consider actions of the type \begin{equation}
S=\int \left (\left[\tilde B_2\wedge \tilde F+\tilde C_3\wedge \CD^{\tilde
A}\tilde \varphi\right]^0_4
+d^4 x\left(F^2_{\mu \nu}+ D_\mu ^2\varphi +V(\varphi )\right) \right)
\end{equation}
 and that after gauge-fixing  and symmetry breaking, one obtains effectively
\begin{equation}
S=\int d^4 x  \left( tr(F^2_{\mu\nu})+\ldots +{\rm supersymmetric\ term}
\right)
\end{equation}
where $ trF^2_{\mu\nu}$ means the trace  of $F^2_{\mu\nu}$
in the broken gauge directions and the supersymmetric terms stand for the ghost
interactions  coming from the gauge-fixing.

A third  more elementary  possibility for  softening the topological invariance
is  to consider a pure $B_2 F$ model coupled to a Higgs field
\be
S_4=\int T_r \left( \left[  \tilde B_2 \wedge F^{\tilde A}\right ]^0_4+ d^4
x\left(
D_\mu \varphi
D^\mu \varphi +V(\varphi)\right)\right) \ee
By symmetry breaking due to the Higgs field, one obtains mass terms for the
Yang-Mills field, and thus a Freedman-Townsend model yielding  a non linear
sigma model  in
the broken directions and a topological  BF  model in the unbroken directions.

\section{conclusion}

We have shown that the B-V  formalism for the gauge theories forms
coupled to   Yang-Mills   forms can be formulated in  a
unifying  algebraic framework. The main idea is to group all relevant fields
and anti-fields for the B-V quantization of a p-form gauge field as  the
components of   differential forms which are graded by the sum of the ghost
number and
ordinary form degree. This suggests that  a p-form gauge field  comes in a
"dual"  pair with a (D-p-1)-form gauge field. In this way, we have obtained an
algoritm which generates  topological  actions function  of such  p-form gauge
fields which are of the Chern-Simon  and/or
Donaldson-Witten type in any given space-time dimension,  on the basis of
vanishing curvature conditions.
 We have indicated that some  of  the  models  which arise  in
this straightforward construction could  undergo a symmetry breaking mechanism.
The latter would soften   the requirement  that all components of the
classical field strenghts vanish classically  and possibly determine  actions
with physical excitations.  In a separate publication, we will show how to
generalize  our observations to the case of 2D reparametrization invariance.

\vskip 1cm
\noindent {\bf{Aknowledgments: }}{The author would like to
express his deep gratitude to RIMS for the hospitality
extended to him during his stay in Japan.}

 \newpage
%


\end{document}